\documentclass[aps,useAMS,usenatbib]{PoS}
\usepackage{natbib}
\usepackage{amsmath}        % need ams stuff    %
\usepackage{amsfonts}       % to make equations %
\usepackage{amssymb} 
\usepackage{color,graphicx}

\title{Fast radio bursts: recent discoveries and future prospects}

\ShortTitle{Fast Radio Bursts}

\author{\speaker{Emily Petroff}\thanks{Speaker.}\\
        ASTRON, The Netherlands Institute for Radio Astronomy, Postbus 2, 7990 AA Dwingeloo, The Netherlands\\
        E-mail: \email{petroff@astron.nl}}

\abstract{Fast radio bursts (FRBs) are quickly becoming a subject of intense interest in time-domain astronomy. The progentiors of FRBs remain unknown but a wide variety of models exist from cataclysmic to repeating scenarios. Advances in FRB detection using current and next-generation radio telescopes will enable the growth of the population in the next few years. Real-time discovery of FRBs is now possible with 6 sources detected in real-time within the past 2 years at the Parkes telescope. Here we discuss the developing strategies for maximising real-time science with FRBs including polarisation capture and multi-wavelength follow-up, with particular focus on real-time detections with the Parkes telescope as a test bed for fast radio burst science. We also discuss how our response to these events can pave the way for the next generation of FRB searches with wide-field interferometers.}

\FullConference{11th INTEGRAL Conference Gamma-Ray Astrophysics in Multi-Wavelength Perspective,\\
		10-14 October 2016\\
		Amsterdam, The Netherlands}

\begin{document}

\section{Introduction}

\begin{figure}
\centering
\includegraphics[width=10cm]{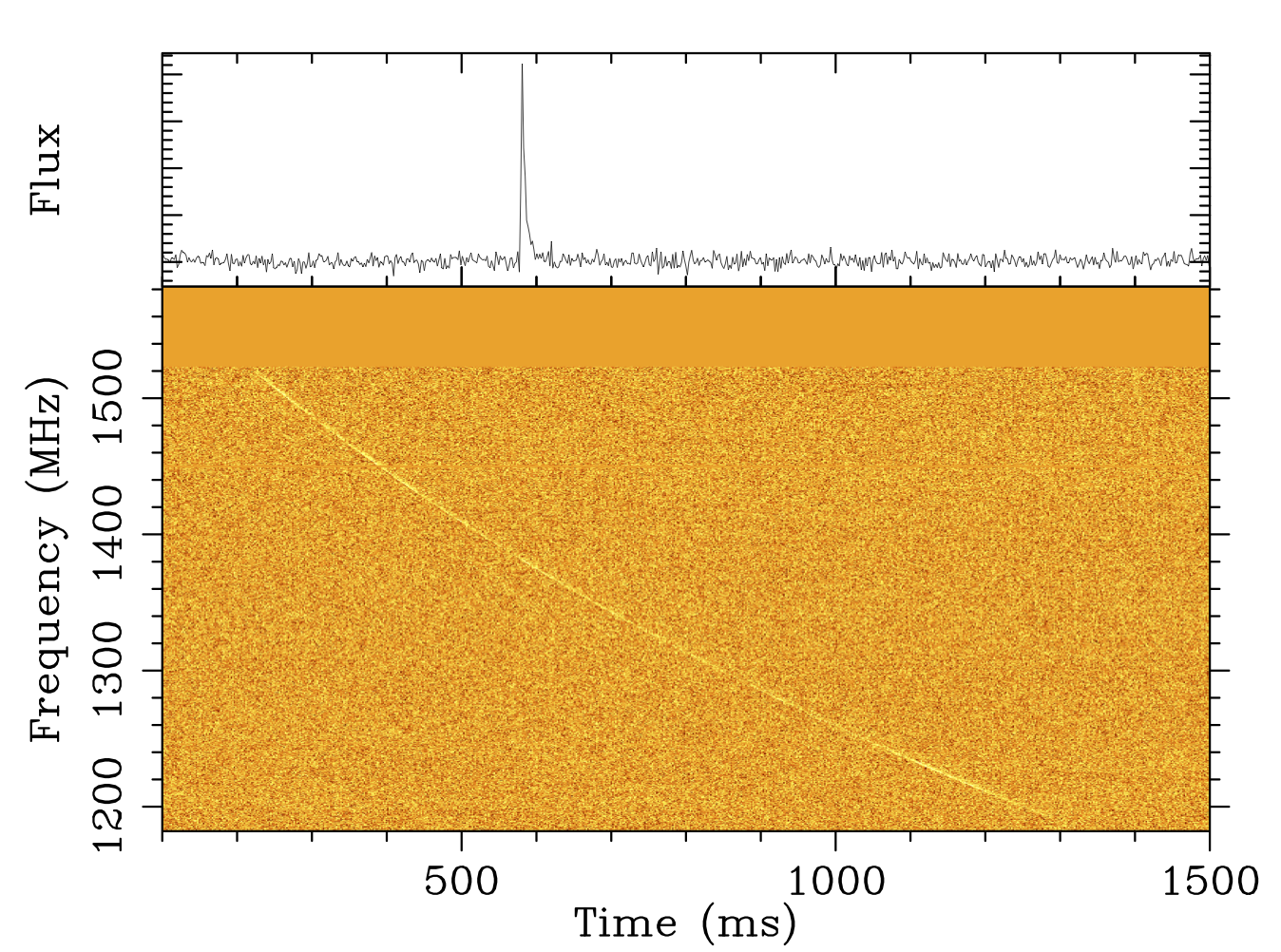}
\caption{The pulse profile (top) and frequency-time spectrum (bottom) of the fast radio burst FRB 110220. The frequency-dependent pulse arrival time is due to dispersion from the ionised electrons along the line of sight. When this effect is removed the pulse duration is $\sim$5 ms. \label{fig:FRB}}
\end{figure}

Fast radio bursts (FRBs) are an emerging class of radio transients observed as bright ($\sim$Jy), short ($\sim$ms) highly dispersed radio pulses \citep{Lorimer07,Thornton13}. FRBs are characterized by an extremely high dispersion measure
\begin{equation}\label{eq:DM}
\mathrm{DM} = \int^{D}_0 n_e d\ell
\end{equation}
\noindent where $n_e$ is the electron column density integrated from the source to an observer at distance $D$. The DM is observed as a quadratic frequency dependent time delay of the pulse arrival time over a radio observing bandwidth, Figure~\ref{fig:FRB}. The DMs of FRBs are many times higher than the column densities expected from Milky Way electrons leading to theories that they originate extragalactically. If a substantial fraction of the DM is due to the intergalactic medium (IGM) this would place the sources of FRBs at cosmological distances \citep{Ioka03,Inoue2004} and make FRBs a unique probe of the diffuse warm ionized intergalactic medium \citep{McQuinn2014}. 

The first FRB was reported in 2007 by \citeauthor{Lorimer07} as a single bright radio pulse discovered in radio survey data from the Parkes telescope recorded in 2001. The `Lorimer burst' remains one of the brightest sources in the FRB population, and the location of the burst was re-observed for hundreds of hours but no repeating bright pulses or periodic signal of any kind were found. 
%\textit{Following the discovery of this bright burst, more Parkes survey data was searched for similar signals and in 2010 (Burke-Spolaor et al) reported the discovery of `perytons' which mimicked the dispersed signal of the Lorimer burst and were found with a similar DM, but were detected at equal intensity in all 13 beams of the Parkes multibeam receiver (cite) suggesting a terrestrial, local origin. The discovery of perytons placed suspicion on the origins of the Lorimer burst and few papers were published on the subject following (Burke-spolaor et al).} 
After several years without a subsequent FRB discovery the field was renewed in 2013 with the publication of four new FRBs by \citet{Thornton13}. Since then, 23 FRB sources have been published in the literature from discoveries at five radio telescopes\footnote{All published FRBs are available in the FRB Catalogue: \texttt{http://www.astronomy.swin.edu.au/pulsar/frbcat/} \citep{frbcat}} \citep{Keane11,SarahFRB,Spitler14,GBTBurst,Ravi2015,PetroffFRB,Keane2016,Champion2016,FRB150807,UTMOSTFRBs,FRB150215,ASKAPFRB}. 
%Additionally, more confidence was placed in the astrophysical origin of FRBs when the source of the perytons was discovered to be the microwave ovens on the Parkes site and it was shown that these microwave ovens could not be responsible for the FRBs discovered at Parkes, including the Lorimer Burst (petroff).

The progenitors of FRBs remain a mystery; however, a substantial number of theories have been proposed in the last five years to explain the varied observational properties of the population. In the most general sense, origin theories for FRBs can be grouped into two categories: cataclysmic and repeating progenitor models. Cataclysmic progenitor theories include collapsing neutron stars \citep[`\textit{blitzars}'; ][]{Falcke}, binary neutron star mergers \citep{Totani2013}, and mergers of charged black holes \citep{Zhang2016}, while non-cataclysmic models include energetic neutron stars in supernova remnants \citep{Connor2016}, giant pulses from young neutron stars \citep{Cordes2016}, and hyperflares from magnetars \citep{Popov2010}. 

Of the 23 FRBs that exist in the published literature none have been conclusively linked to a single progenitor model. One FRB, FRB 121102, has been seen to repeat multiple times, providing a single case where a cataclysmic model has been invalidated \citep{Spitler14,Spitler2016}. In the case of FRB 121102 the detection of multiple pulses allowed for precise localization of the FRB to a dwarf galaxy ($M_* \sim (4-7) \times 10^7 M_\odot$) at a distance of $\sim$1 Gpc, the first precise distance obtained to an FRB and the first concrete proof of a cosmological origin \citep{Chatterjee2017,Tendulkar2017,Marcote2017}. Since no other FRB has been seen to repeat, some even after hundreds of hours of follow-up observations, it remains unclear if FRB 121102 is representative of FRBs in general, is a particularly active or young specimen in the population, or is drawn from a different sub-population than the other FRBs that have not been seen to repeat. The repetition of FRB 121102 is itself unusual as more than a hundred pulses have already been observed but the pulses are highly clustered in time and no pulse period has yet been found (J. Hessels, private communication). Ultimately, more FRBs and more observations of each FRB are needed to fully understand their origins. 

\section{Current Searches and Recent Results}

The priorities of current observing efforts in the field of FRBs remain focused on finding more sources and re-observing the discovery locations to search for repeating pulsed emission. FRB searches are currently ongoing with several facilities around the world including the Upgraded Molonglo Synthesis Telescope \citep[UTMOST; ][]{UTMOSTFRBs}, the Arecibo Telescope \citep{Spitler14}, and the Australia Square Kilometre Array Pathfinder \citep[ASKAP; ][]{ASKAPFRB}. Here we focus on the recent efforts and results from the Parkes radio telescope, in particular the FRB searches performed as part of the Survey for Pulsars and Extragalactic Radio Bursts \citep[SUPERB; ][]{SUPERB}. 

Searches for FRBs with the Parkes telescope have been particularly fruitful due to the rapid survey speed facilitated by the 13-beam multibeam receiver. This coupled with the high telescope sensitivity and the high fractional available observing time for pulsar and FRB searches has made Parkes the most successful telescope in the world currently searching for FRBs. The SUPERB survey is currently the primary pulsar and FRB survey running at Parkes; its goal is to search the Southern sky ($\delta < +10^\circ$) at intermediate and high Galactic latitudes ($|b| > 15^\circ$) in real-time for bright single pulses and periodic pulsar signals \citep{SUPERB}. 

The real-time FRB search is performed using the \textsc{Heimdall} single pulse search software\footnote{http://sourceforge.net/projects/heimdall-astro/} which is capable of processing incoming data at $\sim1.5\times$ real time and sending a notification upon detection of an FRB-like signal. The advantage of real-time detection capability is two-fold: a real-time detection with the Parkes system enables the polarization information of the burst to be preserved, and it also allows for the triggering of rapid multi-wavelength follow-up to search for associated emission. The real-time search system began operation in March 2014 and since its deployment six FRBs have been discovered in real-time; in each case, the burst polarization information was preserved and in most cases multi-wavelength follow-up was initiated within hours of the FRB discovery \citep[][Bhandari et al., in prep.]{PetroffFRB,Ravi2015,Keane2016,FRB150215}. In the following subsections we summarize some of the most recent and promising results from these efforts.

\subsection{FRB Polarization}

Only five FRBs have published polarization information but already the study of polarization properties for these bursts is of great interest within the community. Polarization profiles with a substantial fraction of linear and/or circular polarization provide insights into the emission mechanism at the source and the alignment of the charged particles responsible for the coherent radio emission. Of particular interest, however, is the measurement of Faraday rotation due to magnetic fields along the line of sight given by the rotation measure

\begin{equation}
\mathrm{RM} = \int^{D}_0 n_e B_\parallel d\ell
\end{equation}

\noindent where $D$ and $n_e$ are identical to Equation~\ref{eq:DM}, and $B_\parallel$ is the magnetic field parallel to the line of sight. Although this is an integrated quantity along the entire line of sight the measured value is dominated by the contributions from the host galaxy and the Milky Way foreground, as the magnetic field strengths in the IGM are many orders of magnitude weaker those those in galaxies. In many cases the Milky Way foreground can be estimated from surveys of polarized extragalactic sources \citep{Taylor2009}, making it potentially possible to directly measure magnetic fields local to the FRB progenitor and in the host galaxy. 

Three published FRBs have robust RM measurements, two from Parkes: FRB 150807 and FRB 150215 \citep{FRB150807,FRB150215}, and one from the Green Bank Telescope: FRB 110523 \citep{GBTBurst}. Two other published FRBs have available polarization data but were both found to have insufficient linear polarization to make an RM measurement: FRB 140514 and FRB 150418 \citep{PetroffFRB,Keane2016}. Somewhat frustratingly, no coherent picture has yet emerged from the small sample of available RMs. FRBs 150807 and 150215 were found to have RMs of 12.0$\pm$0.7 rad m$^{-2}$ and 1.5$\pm$10.5 rad m$^{-2}$, respectively, both consistent with the Milky Way foreground at their respective sky locations. Conversely, FRB 110523 was found to have RM = -186.1$\pm$1.4, much greater than the contribution expected from the Galactic foreground. An agreement with the Galactic foreground RM indicates no integrated magnetic field contribution to the RM from the FRB host galaxy, whereas an RM excess indicates an ordered magnetic field contribution from the FRB host galaxy. 

The limited polarization data available for FRBs makes it difficult to provide a physical explanation for these differences at present. The host galaxy and progenitor region could have low magnetic fields in the cases of FRB 150215 and FRB 150807. Alternatively, the absence of an ordered magnetic field contribution from the host galaxy could be caused by a magnetized but highly turbulent progenitor region where disordered magnetic fields contribute no net RM. While the prospect of directly probing the magnetic fields of host galaxies billions of lightyears away is quite tantalizing, significant development is needed in the measurements of not only the RMs of FRBs but also the RM contributions from the Milky Way foreground. Insufficient foreground coverage has already proved to be an issue in some cases \citep{FRB150215} and will continue to be valuable information as more FRB RMs are measured. Early results from these three FRBs already suggest that this will be an area of continued interest as the sample of polarized FRBs grows.

\subsection{Multi-wavelength follow-up}

The identification of an FRB event in the seconds or minutes after it is detected by a radio telescope can also enable the triggering of multi-wavelength follow-up. It is currently unknown whether the progenitors responsible for FRBs also produce signatures at other wavelengths. Many repeating progenitor theories such as magnetar and young pulsar models do not predict any associated emission at other wavelengths \citep{Cordes2016,Metzger2017} as the radio engine would be similar to that of a pulsar, which is typically confined to the radio regime. Cataclysmic models may produce multi-wavelength emission, for example in the form of associated prompt gamma-ray emission with short-lived X-ray and optical afterglows, in the case where FRBs are produced in binary neutron star mergers, similar to short GRBs \citep{Totani2013}. Other cataclysmic progenitor models that predict a short-duration explosion followed by the outward propagation of an ionized or irradiated shell also predict transient optical or radio emission from the source on timescales of days to months \citep{Zhang2016}. 

The search for associated emission from FRBs is currently complicated by the fact that single dish telescopes such as Parkes and Arecibo have large beams on the sky in which sources cannot be more precisely localized (the Parkes beam full-width half-maximum (FWHM) is 14.4$'$). The number of galaxies in such a large area on the sky prohibits unambiguous association from the radio pulse alone, therefore the current strategy to search for associated emission is to detect the FRB in real-time and search for variable or transient emission post-burst that can be related back to the FRB. 

Searches for these multi-wavelength signatures have been initiated for four published FRBs from the Parkes telescope: FRBs 131104, 140514, 150215, and 150418 \citep{FRB131104Radio,PetroffFRB,FRB150215,Keane2016}. In the case of FRB 131104 only radio imaging data was collected in the days after the FRB to search for transient GHz radio emission, but for the other three an extensive multi-wavelength campaign was conducted from X-ray to radio wavelengths. Follow-up for one FRB, FRB 150418, detected a potentially associated 6-day radio transient in the field \cite{Keane2016}, although a subsequent re-brightening of the source in question led to a revised conclusion of AGN variability as the cause of the transient \citep{Williams2016}; an association with the FRB is still not fully ruled out but no conclusive association was possible \citep{Johnston2017}. Despite the extensive searches in these observing campaigns, no transient or variable sources that could be linked back to the FRB sources were detected above the sensitivity thresholds of the telescopes involved.

%The most extensive and prompt multi-wavelength follow-up to-date was done for FRB 150215 \citep{FRB150215}. In this case, 11 telescopes were employed in the weeks and months after the detection of the FRB, from the H.E.S.S. telescope to search for high energy gamma-rays to the Giant Metrewave Radio Telescope (GMRT) at radio frequencies of 610 MHz; a search was also performed for high energy neutrinos using the ANTARES neutrino detector. In all cases the data were searched for variable and transient sources between each observing epoch as well as for variability within each epoch where possible. 

The published observations only probe a limited region of the multi-wavelength transient parameter space. The lack of observations immediately before and after the detection of the FRB, and insufficiently deep observations per epoch may have contributed to the limited of success of these searches. For example, some predict that FRBs may be associated with kilonovae \citep{Niino14} in which case the optical emission, which would be several magnitudes fainter than current limits in published searches, would not be detected. In a more extreme optical precursor example \citeauthor{Metzger2017} propose that FRBs are generated by young magnetars created by superluminous supernovae (SLSNe). In this case, the SLSN would occur decades to centuries before the FRB emission is seen and an association for the FRBs in the current sample is difficult with the incompleteness of archival supernova surveys. Ultimately, a larger population of FRBs is needed with both commensal and rapid follow-up observations ($<$1 hour post-burst) at X-ray, optical, and radio wavelengths followed by high cadence observations to adequately sample the lightcurve of any possible transient. More precise localization of FRBs from their detection pulses will also provide a much smaller error circle, thereby dramatically reducing the number of possible sources in the error region; this will allow more telescopes with higher sensitivity and smaller field of view to be used in future follow-up endeavors. 

\section{Future Searches}

\begin{figure}
\centering
\includegraphics[width=14cm]{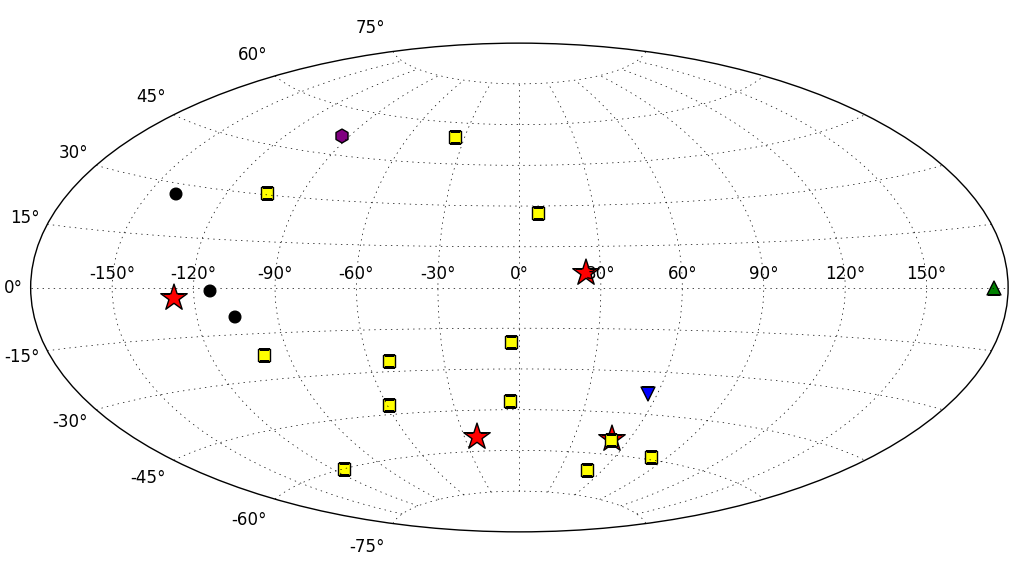}
\caption{An aitoff projection of the Galactic latitude and longitude positions of all published FRBs showing Parkes archival FRBs (yellow squares), Parkes real-time FRBs (red stars), the Arecibo FRB (green triangle), the GBT FRB (blue inverted triangle), the UTMOST FRBs (black circles), and the ASKAP FRB (purple hexagon).  \label{fig:aitoff}}
\end{figure}

The rate of progress in the field of FRB research is increasing at an impressive rate. Within the past five years the field has seen a wealth of discoveries and a host of theoretical papers around the subject. Currently only 23 FRB sources are published in the literature, Figure~\ref{fig:aitoff}, but this number is expected to increase as ongoing surveys continue their searches and as new instruments come on-line to search the sky with a larger field of view and at a higher cadence. The majority of FRBs have been found with large collecting area single dish radio telescopes; however, enormous gains have already been achieved using wide-field radio interferometers which have the critical power of better localizing FRBs to identify a host galaxy \citep{Chatterjee2017}. The UTMOST interferometer in Australia has already detected three FRBs \citep{UTMOSTFRBs} and, more recently, the ASKAP interferometer detected one FRB in a 3.4-day pilot survey \citep{ASKAPFRB}. 

The next generation of interferometric radio dish arrays such as ASKAP in Australia and the Aperture Tile In Focus \citep[Apertif; ][]{Apertif} upgrade to the Westerbork Telescope in the Netherlands are equipped with phased array feeds (PAFs), multi-element receivers capable of forming multiple beams on the sky over a large field of view. The new PAF technology combined with the interferometric capability to more precisely localize an astronomical source on the sky makes these new instruments ideal machines for finding FRBs. Initial results from ASKAP suggest that these arrays may find as many as a few FRBs per week. Similarly, cylindrical arrays such as UTMOST in Australia and the Canadian Hydrogen Intensity Mapping Experiment \citep[CHIME; ][]{CHIMEFRB} telescope in Canada which operate in a drift scan mode observe an enormous field of view ($\sim 250^{\circ}$ for CHIME) and are expected to find hundreds of FRBs per year. Crucially, they will also be able to re-observe the location of the FRB as it drifts overhead every day.

The combination of upcoming surveys with these powerful new instruments and proposed search efforts at existing telescopes augur a population explosion in the next decade. The expected FRB discovery rate from Apertif alone is expected to more than double the current population within a year of operation at design sensitivity \citep{ARTS2014}. Thus, the future is incredibly bright for FRB science. A large number of unanswered questions remain as to the origins, population properties, emission features, repeatability, and multi-wavelength emission of FRBs, but the expected population boom in the coming years may provide some valuable and interesting answers.

\section*{Acknowledgments}

\noindent Funding from the European Research Council under the European Union's Seventh Framework Programme (FP/2007-2013) / ERC Grant Agreement n. 617199. Parts of this research were conducted by the Australian Research Council Centre of Excellence for All-sky Astrophysics (CAASTRO), through project number CE110001020 and the ARC Laureate Fellowship project FL150100148.

\end{document}